\documentclass[onecolumn]{aa}  

\usepackage{graphicx}
\usepackage{txfonts}
\usepackage{lipsum}
\usepackage{subcaption}        
\usepackage{lscape}            
\usepackage{placeins}

\begin{document}

   \title{Rebrightenings of gamma-ray burst afterglows from an increasing magnetic inclination angle of a nascent magnetar}
   \titlerunning{Rebrightenings from an increasing magnetic inclination angle}
   \authorrunning{Xu et al.}

   \author{M. Xu\inst{1}\fnmsep\thanks{Corresponding author:xuming@nju.edu.cn}, J. Li\inst{1}, C. F. Xiao\inst{1} \and H. H. Qiu \inst{1}
        }

   \institute{Department of Physics, Jiangxi Science and Technology Normal University, 330013, Nanchang, China\\
             }

   \date{Received 00 00, 20XX}

  \abstract
   { A nascent magnetar, accompanying a gamma-ray burst (GRB) explosion, releases enormous rotational energy via magnetic dipole radiation.
     The energy loss rate of the magnetar is determined by the strength of the magnetic field at the pole.}
   { We investigated the effect of the magnetic inclination angle on the energy loss rate. The released energy is injected into the GRB jet
     and shapes the light curves of GRB afterglow. Different evolutionary approaches lead to different curves shapes.}
   {A shallow decay phase in GRB X-ray afterglow may result from energy injection from a magnetar with a fixed inclination angle.
     A two-plateau phase may result from a decreasing inclination angle scenario.
     In this study, we considered an increasing inclination angle scenario.
     The energy loss rate of the magnetar increases as the magnetic inclination angle grows.
     }
   { Our analysis reveals that as the lost rotational energy injected into the GRB jet increases,
       rebrightening phases occur in the GRB afterglows.
      The rebrightening features are slight and short-lived.}
   { The observed afterglow rebrightening of GRB 170822A and GRB 230414B can be well explained within our framework.
     Some GRB X-ray afterglows that exhibit slight and early rebrightenings may result from an
     increasing magnetic inclination angle of a nascent magnetar.}


   \maketitle
\titlerunning{GRB} \authorrunning{Xu}

\section{Introduction}

Gamma-ray bursts (GRBs) are among the most powerful electromagnetic explosions in the Universe.
These events are empirically classified into long and short durations,
historically linked to collapsars and compact binary mergers, respectively
\citep{Blin84,Eich89,Woos93,Piran99,Zhang07,Zhang18}.

The prompt and afterglow emissions of both long and short GRBs can originate from nascent
magnetars accompanying the bursts
\citep{Usov92,Dai98,Viet98,Zhang01,Dall11,Rowl13,Lasky16,Grei15,Fu24,Roman25}.

A nascent magnetar is a highly magnetized neutron star with a short spin period.
When a magnetar is born, it spins down and loses energy through multiple channels.
At a very early stage, a strong magnetic wind carries away part of its angular momentum
\citep{Metz11,Land20}.
Gravitational wave emission plays an important role if the magnetic deformation is sufficiently large
\citep{Cutl01,Cutl02,Dall09}.
A substantial amount of rotational energy is extracted through magnetic dipole radiation
\citep{Paci67,Gold69,Spit06}. Additional losses may occur through higher-order magnetic
multipoles and other complex structures.
\citep{Petri15,Wang24,Rayn20}.

For the magnetic dipole radiation model,
the energy released from a magnetar is injected into the GRB jet outflows.
If the injected energy is comparable to the radiation energy of the GRB jet,
a shallow decay (or plateau) phase appears in the afterglow light curve
\citep{Fan06,Xu09,Li12,Zhang07,Zhang18,Zhong24}.

Some GRB afterglow light curves exhibit rebrightening (or bump) features
\citep{Liang13,Yang24}.
This phenomenon can be explained by either a 
structured GRB jet geometry
\citep{Huang04,Xu10,Asce20,Beni20,Ogan20}
or through late-time energy injection from a new component
\citep{Deng10,Yu15,Ren22,Geng25}.

In this paper, we present a new model to explain the rebrightening structure
without requiring a special jet or a new injected component. In our model, a magnetar spins down
with an increasing magnetic inclination angle. The released energy increases through
magnetic dipole radiation and is injected into the GRB outflows.
A rebrightening structure naturally appears in the afterglow light curves.

Our paper is organized as follows.
Section 2 describes our model. Section 3 presents our numerical results.
In this section, we present our fit result to the observed data.
Section 4 provides a discussion and conclusions.

\section {Jet dynamics and energy injection}

A collimated ultra-relativistic jet is launched from the central engine during a GRB event.
When this jet interacts with the interstellar medium (ISM), it produces shocks.
The afterglow emission arises primarily from external shocks. When the central engine is a
magnetar, the energy loss from the magnetar spinning down injects additional power into the shocks.
This energy injection can reshape the afterglow light curve,
producing characteristic features such as rebrightenings.

\subsection {Jet dynamics}
The dynamical evolution of a beamed outflow has been comprehensively modeled by Huang et al.
\citep{Huang99,Huang00,Huang03}.
The bulk Lorentz factor ($\gamma$) of the outflow evolves as follows:
\begin{equation}
  \frac{d\gamma}{dt}=\frac{-(\gamma^{2}-1)}{{M_{\rm{ej}}+\epsilon m+2(1-\epsilon)\gamma m}}\times
  \frac{{\rm d}m}{\rm{d}t},
\end{equation}
where $M_{\rm ej}$ is the initial ejecta mass, $m$ is the
swept-up mass, and $\epsilon$ is the radiation efficiency.

When energy injection is included, the differential equation is modified to
\citep{Xu09,Kong10,Xu21}
\begin{equation}
\label{equ:g}
  \frac{d\gamma}{dt}=\frac{1}{{M_{\rm{ej}}+\epsilon m+2(1-\epsilon)\gamma m}}\times
  [\frac{\eta L}{c^{2}}-(\gamma^{2}-1)\frac{{\rm d}m}{\rm{d}t}],
\end{equation}
where $L$ is the energy injection luminosity and $\eta=\frac{1}{2}(1-{\rm cos} \theta)$
($\theta$ is the half-opening angle of the GRB jet) is the fraction of $L$ injected into the collimated outflow.

\subsection {Energy loss rate of a magnetar}
The energy injection luminosity term in Eq. (\ref{equ:g}) may originate from a nascent magnetar
via magnetic dipole radiation.
The proto-neutron star is likely surrounded by a plasma-filled magnetosphere
\citep{Gold69}
rather than by a vacuum
\citep{Paci67}.

During the initial spin-down phase, the magnetar's dipole radiation luminosity evolves
as
\citep{Spit06}
\begin{equation}
\label{equ:L}
   L_{\rm{dip}}=\frac{B_{\rm{p}}^{2}R^{6}\Omega^{4}(1+\rm{sin}^{2}\alpha)}{6 c^{3}}.
\end{equation}
The luminosity $L_{\rm{dip}} = I\Omega\dot{\Omega}$ quantifies the magnetar's rotational energy loss rate.
The parameters include $c$ (speed of light), $B_{\rm p}$ (polar magnetic field strength), $R$ (stellar radius),
$\Omega$ (angular velocity), $\dot{\Omega} \equiv d\Omega/dt$ (spin-down rate), $\alpha$ (magnetic inclination angle),
and $I$ (moment of inertia).

During the process of energy loss, the magnetic inclination angle ($\alpha$) is very important.
Its evolution determines the energy loss rate of the magnetar.

A newborn magnetar may have an ellipsoid shape and be unstable, with its magnetic axis misaligned with the rotational axis
\citep{Davis70,Mich70,Mest81}.
In a spin-down scenario with a fixed inclination angle, the energy loss rate exhibits a shallow decay structure
\citep{Xu09}.

As the star cools and spins down, its interior may undergo bulk viscosity while externally
experiencing electromagnetic torque
\citep{Mest72,Jones75,Jones76}.
For an oblate star, both the bulk viscosity and external electromagnetic emission tend to decrease the inclination angle.
A two-plateau energy loss rate may result from this decreasing inclination angle scenario
\citep{Xu21}.

In this work, we assume that the star is a prolate ellipsoid whose deformation originates from a strong internal toroidal magnetic field.
The magnetic inclination angle increases rapidly toward $90^\circ$ due to viscous damping
\citep{Jones75,Jones76,Land17,Land18,Land20}. 

A prolate ellipsoid is an efficient emitter of gravitational waves
\citep{Cutl01,Cutl02,Dall09}.
The spin-down of the star results from both electromagnetic and gravitational wave emission.
for our model (using the parameters in Table 1), 
the magnetic ellipticity is approximately $10^{-7}$, and the corresponding gravitational wave luminosity is about
$10^{40}\; \rm{ erg\cdot s^{-1}}$ 
\citep{Cutl01,Dall09}
, while the electromagnetic emission luminosity is about $ 10^{49} \;\rm erg\cdot s^{-1}$.
The electromagnetic emission is far larger than the gravitational wave emission.
Therefore, within our scenario, the gravitational wave contribution is small and can be neglected.

We used and simplified the inclination angle evolution model from
\citet{Land18}.
As shown in their Figure 8, the inclination angle remains nearly constant at an early stage, then increases linearly
to the termination angle, and finally stabilizes. Our model assumes an evolution timescale $T_0$: the inclination
angle stays constant for $T_0/3$, then increases linearly from the initial value to the termination value over the
remaining $2T_0/3$. We set the initial angle to 0.01 rad and the termination angle to $\pi/2$ rad.

The evolution of the increasing angle of inclination is as follows:
\begin{equation}
\label{equ:a}
\alpha=
\left\{
  \begin{array}{lr}
 \qquad \quad \alpha_{\rm i}, \qquad \qquad \qquad  t\leq T_0/3, &  \\
  &  \\
  \alpha_{\rm i}+\frac{(\alpha_{\rm t}-\alpha_{\rm i})\cdot (t-T_0/3)}{2T_{0}/3}, \qquad   T_0/3< t \leq T_0, &  \\
  &  \\
  \qquad \quad \alpha_{\rm t},  \qquad \quad \quad  \qquad   \;\ t> T_0. &
  \end{array}
\right.
\end{equation}
The angle increases rapidly from the initial inclination angle $\alpha_{\rm i}$
to the termination angle $\alpha_{\rm t}$ on a timescale of $T_0$.

The timescales $T_0$ used in our model (see Table 1) are several hundred to several thousand seconds.
These timescales are larger than the bulk viscosity timescale reported in
\citet{Land18}.
The timescale formula is parameter-dependent.
For our parameters (see Table 1), if we take the proto-neutron star temperature (parameter T in Table 1)
to be $1.44\times10^{10}$ K for GRB 170822A and $1.8\times10^{10}$ K for GRB 230414B (see Equation 59 in Lander \& Jones 2018),
we can recover the timescales.

\subsection {Energy injection luminosity}

Interaction of the ultra-relativistic GRB ejecta with the interstellar medium produces an external
forward shock. Electrons are accelerated by the shock and produce afterglow emission
through synchrotron radiation. In the energy injection model, the spin-down Poynting-flux energy refreshes
the forward shock, slowing the synchrotron cooling of the electrons.

In our model, the energy injection luminosity is proportional to the magnetar's energy loss rate (i.e., $L=L_{\rm dip}$).
We injected a factor $\eta$ of this power into the outflows.
If the luminosity is sufficiently higher than the emission from the
shock-accelerated electrons, a rebrightening phase appears in the afterglow light curve.

Using certain magnetar parameters described in Section 3.1, we calculated the energy injection luminosity. As shown by the solid curve in Fig. \ref{fig1-ELR},
as the magnetic inclination angle increases, the energy injection luminosity
(i.e., the energy loss rate) also increases.
It varies by a factor of approximately two, consistent with the limit imposed by Eq. (\ref{equ:L}).

\subsection {Energy injection and afterglow emission}

The increasing energy is continuously injected into the external forward shock,
which results from the interaction between the GRB jet outflow and the ISM.
The jet dynamical evolution follows Eq. (\ref{equ:g}).
The afterglow emission arises from synchrotron radiation produced by
relativistic electrons within the jet 
\citep{Pacz93,Mesz97,Huang99,Huang00}.

When this injected energy becomes comparable to the synchrotron radiation energy
of shock-accelerated electrons,
it significantly reshapes the multiband GRB afterglow light curves.
These modified curves exhibit distinct features relative to the no-injection scenario.
As the energy injection luminosity ($L$) increases, a rebrightening phase emerges in the afterglow light curves.

\section {Numerical results}

In our previous studies \citep{Xu09,Xu15,Xu21}, we extensively studied models of fixed or decreasing magnetic inclination angles.

In contrast, the present work focuses on a model in which the magnetic inclination angle increases over time.
This temporal increase in the inclination angle leads to a corresponding increase in the injected luminosity.
We numerically calculated the full dynamical evolution of the GRB outflow under this scenario and present the detailed results below.

\subsection {Energy loss rate}

In our calculations, we adopt the parameters below for the proto-neutron star, consistent with those used in our fit to GRB 230414B (see Table 1).

\begin{table*}
\label{tab:paras}
\begin{center}
{
\begin{tabular}{|c|c|c|c|c|c|}
\hline
 parameters& 170822A & 230414B &parameters & 170822A & 230414B \\
\hline
R/km          & 12                &  12                &$z$           &            1.0   &   3.568          \\
$M/M_\odot$   & 1.7               &   1.7              &$E_0$/ergs    &$3.5\times10^{52}$&$6.0\times10^{52}$\\
$B_{\rm p}$/G &$5.0\times10^{14}$ & $2.0\times10^{14}$ &$\theta$/rad  & 0.15             & 0.1   \\
$P_0$/ms      &1.2                &0.75                &$\gamma$      & 80               & 120   \\
$T/K$         &$1.44\times10^{10}$&$1.8\times10^{10}$  &$n_{\rm ISM}$/(cm$^{-3}$)& 1.2   & 0.1    \\
$\alpha_{\rm i}$/rad &  0.01      &0.01                &$p$           &2.9               & 2.6    \\
$\alpha_{\rm t}$/rad & $\pi/2$    & $\pi/2$            &$\varepsilon_{\rm e}$ & 0.16     & 0.1   \\
$T_0$/s       & $7.0\times10^{2}$ & $2.5\times10^{3}$  &$\varepsilon_{\rm B}$ & $0.075$  & $0.03$\\
\hline
\end{tabular}
}
\end{center}
\caption{Model parameters derived from our best fit to GRB 170822A and GRB 230414B.}
\end{table*}

The nascent magnetar parameters adopted in our calculations are as follows: 
The magnetar radius is $R=12.0$ km, the mass is $M=1.7 \;M_\odot$, and the corresponding moment of inertia is $I=2.0\times 10^{45}\; \rm{g\cdot cm^{2}}$.
The surface field strength at the magnetic pole is $B_{\rm p}=2.0\times10^{14}$ G.
The initial spin period is $P_0=0.75$ ms (corresponding to an initial angular velocity of approximately
$\Omega_0 = 8.4\times10^3 \; \rm{rad\cdot s^{-1}}$).
The temperature is $T=1.8\times10^{10}$ K and the corresponding timescale is $T_0=2.5\times10^3$ s.
The magnetic inclination angle is assumed to increase from an initial angle of $\alpha_{\rm{i}}=0.01$ rad
to the termination angle of $\alpha_{\rm t}=\pi/2$ rad over a timescale of $T_0$.

\begin{figure}
\centering
\includegraphics[scale=0.45]{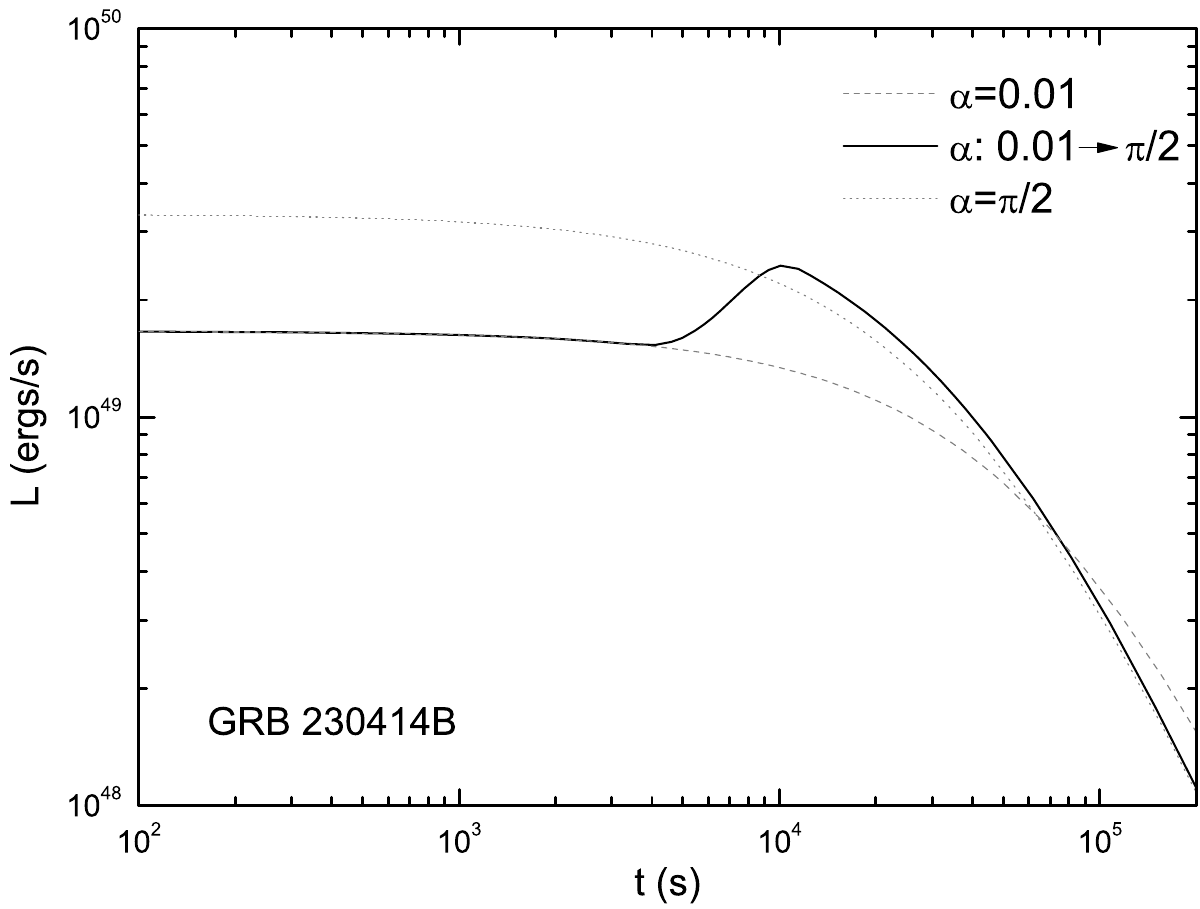}
\caption{\label{fig1-ELR}
Energy loss rate (energy injection luminosity) from nascent magnetar spin-down.
Solid curve: evolving magnetic inclination angle ($\alpha$) from $\alpha_{\rm i} = 0.01$ rad to $\alpha_{\rm t} = \pi/2$ rad.
Dashed gray curve: fixed inclination $\alpha \equiv \alpha_{\rm i} = 0.01$ rad.
Dotted gray curve: fixed inclination $\alpha \equiv \alpha_{\rm t} = \pi/2$ rad.
}
\end{figure}

As the magnetic inclination angle increases, the energy injection luminosity ($L$) exhibits
a significant evolution (solid curve in Fig. \ref{fig1-ELR}).
For comparison, we show two fixed-angle cases: $\alpha=0.01$ rad (dashed gray curve) and $\alpha=\pi/2$ rad (dotted gray curve).

As illustrated by the solid curve in Fig. \ref{fig1-ELR},
the initial magnetic inclination angle of our evolutionary angle scenario
is $\alpha_{\rm{i}}=0.01$ rad. The energy loss rate begins at a low level, similar to that of the fixed-angle case of 0.01-rad (dashed curve).
As the inclination angle increases, the energy loss increases, producing an obvious bump. 
Finally, the magnetic inclination angle stops increasing at $\alpha_{\rm t}= \rm \pi/2$ rad.
The energy loss rate then decreases, approaching the fixed-angle case of the $\pi/2$ rad scenario (dotted curve).
It varies by a factor of approximately two, consistent with the limit imposed by Eq. (\ref{equ:L}).

\subsection {Fits to GRB 170822A and GRB 230414B}

Using the model described above, we calculated the light curves for GRB 170822A and GRB 230414B.
For GRB 170822A
\citep{Page17,Lien17},
no optical counterpart has been observed,
and the redshift is unknown; we assume a value of 1.
We used GRB 230414B as an example to describe the parameters.

For GRB 230414B, we adopted the following outflow parameters (see also Table 1).
The initial energy per solid angle is $E_0=6.0\times10^{52}/4\pi$ ergs
\citep{DAi23,Pars23},
the redshift is $z=3.568$
\citep{Agui23},
the initial Lorentz factor is $\gamma_0 = 120.0$ and the number density of the
ISM is $n_{\rm ISM }=0.1$ cm$^{-3}$.
The power-law index of the electron energy distribution is $p=2.6$,
the electron energy fraction is $\varepsilon_{\rm e}=0.1$, the magnetic energy
fraction is $\varepsilon_{\rm B}=0.03$, and the initial half opening angle of the jet $\theta_0 = 0.1$ rad.
These parameters satisfy the requirements of the external forward shock model.

We also assume a standard cosmology model with $\Omega_{\rm M}$ = 0.315,
$\Omega_{\rm \Lambda}$ = 0.685 and $H_0 = 67.4\;\rm km\cdot s^{-1}\cdot Mpc^{-1}$
\citep{Agha20}.

\begin{figure*}
\includegraphics[scale=0.4]{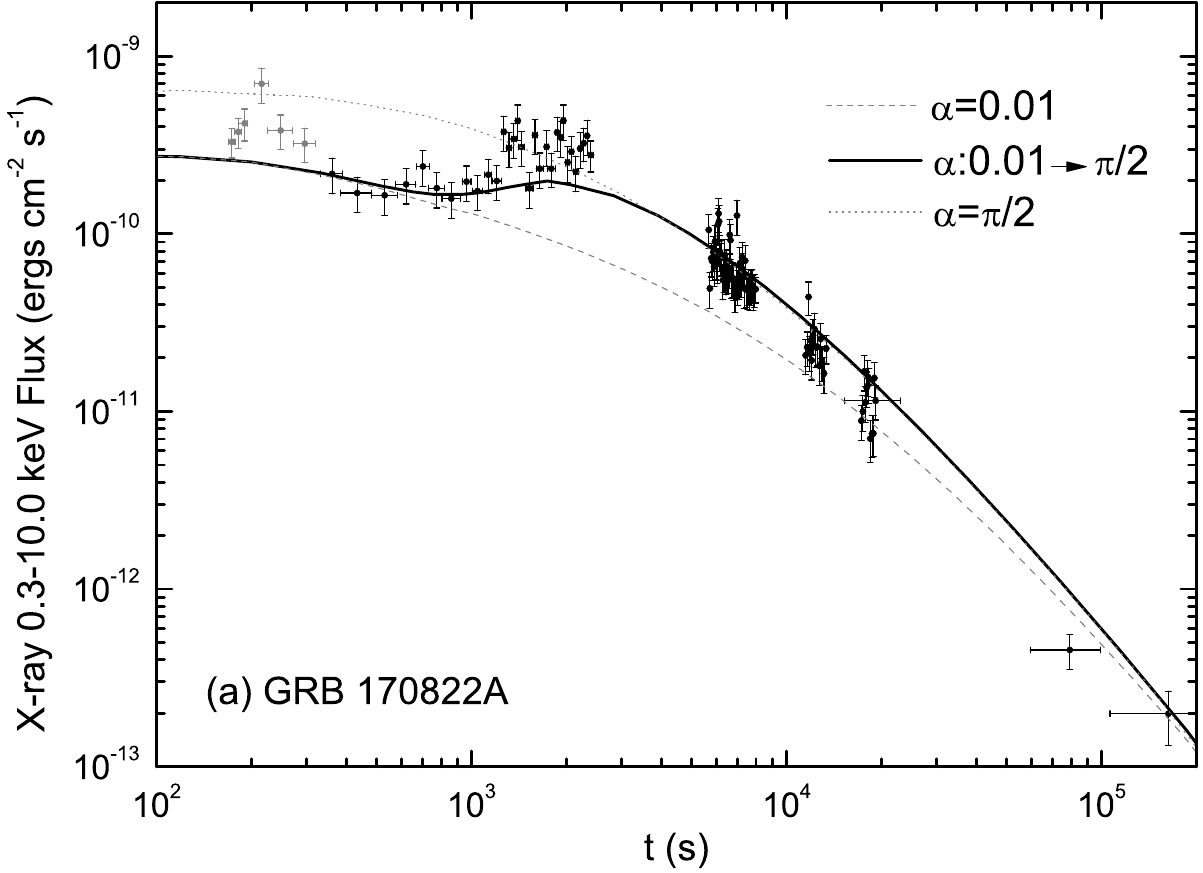}
\includegraphics[scale=0.4]{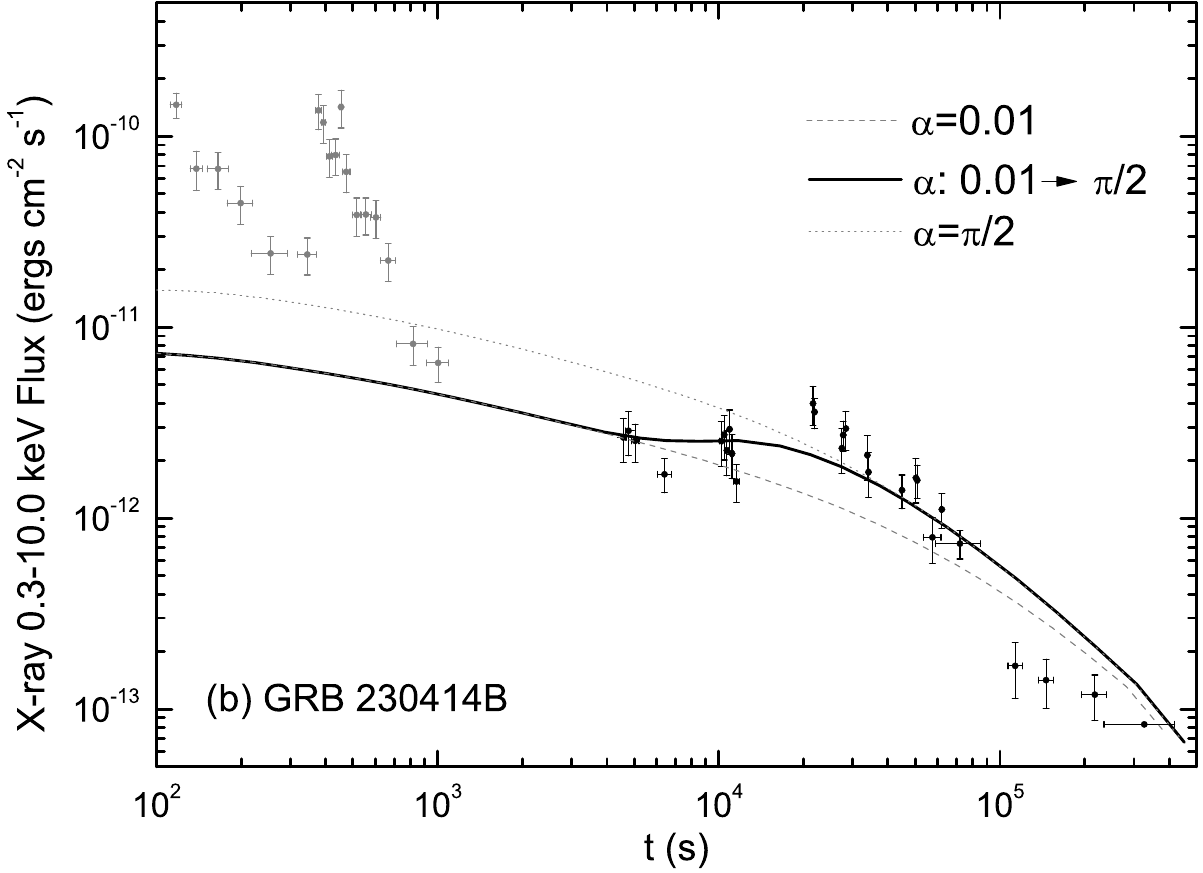}
\caption{\label{fig2-XA}
Observed X-ray afterglows of GRB 170822A (left) and GRB 230414B (right) with theoretical fits.
Data points: Swift-XRT observations.
Solid curve: magnetar model with an evolving magnetic inclination angle
($\alpha$ increasing from $\alpha_{\rm i} = 0.01$ rad to $\alpha_{\rm t} = \pi/2$ rad).
Dashed gray curve: fixed $\alpha \equiv \alpha_{\rm i} = 0.01$ rad.
Dotted gray curve: fixed $\alpha \equiv \alpha_{\rm t} = \pi/2$ rad.
}
\end{figure*}

\begin{figure}
\centering
\includegraphics[scale=0.4]{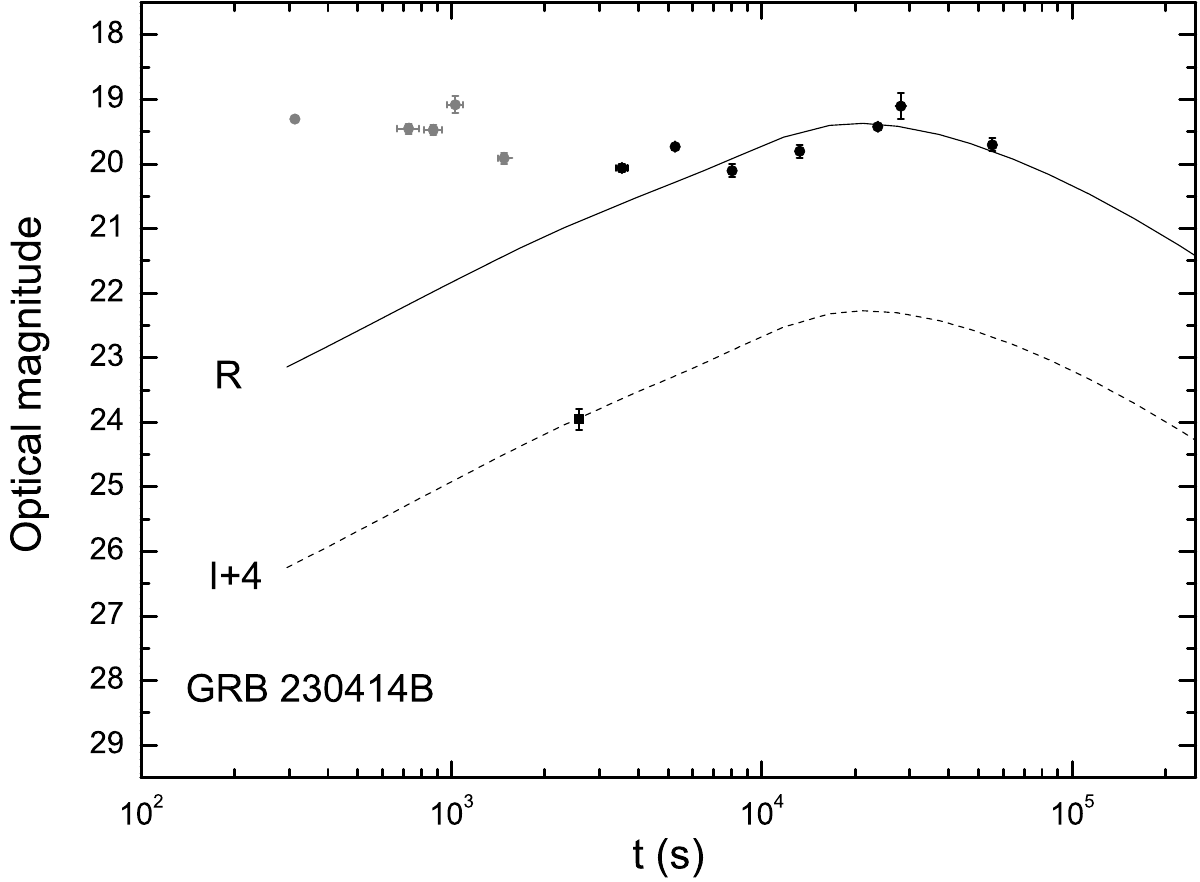}
\caption{\label{fig3-OA}
Optical afterglow light curves of GRB 230414B.
Observational data (GCN circulars): circles and squares correspond to R-band and I-band (add 4 mag) photometry.
Modeled light curves: solid and dashed curves correspond to R-band and I-band (add 4 mag) synthetic flux, respectively
 }
\end{figure}

Using this framework and the parameters in Table 1, we numerically computed the full dynamical evolution
and generated multiband afterglow light curves.
Figures \ref{fig2-XA} (a) and \ref{fig2-XA} (b) show the 0.3-10 keV
X-ray light curves for GRB 170822A and GRB 230414B,
while Fig. \ref{fig3-OA} presents the optical light curves for GRB 230414B.

The X-ray light curves (Fig. \ref{fig2-XA}) incorporate observational data from the Swift satellite
\citep{Gehr04,Bart05,Evans07,Evans09}.
Early-time flares (gray points) originate from ongoing central engine activity.
Following this phase, the external forward shock dominates the emission, producing a smoothly decaying flux profile.
A distinct rebrightening feature is evident in the light curves.
The rebrightening phases in both GRBs are modest, consistent with the constraint imposed by Eq. (\ref{equ:L}).
Subsequently, the light curves resume a normal power-law decay.

In Fig. \ref{fig2-XA}, the solid curve corresponds to the scenario with
our evolving magnetic inclination angle, i.e., $\alpha$ increases from 0.01 rad to $\pi/2$ rad.
The dashed and dotted gray curves correspond to the fixed-angle scenario, with
$\alpha =0.01$ rad and  $\alpha =\pi/2$ rad, respectively.
As shown in Fig. \ref{fig2-XA}, our evolving-angle model provides a good fit to
the Swift-XRT observations of both GRB 170822A and GRB 230414B.

At early times, the solid curve is close to the dashed curve.
The energy injection luminosity is as low as in the fixed 0.01-rad scenario.
Our model is consistent with the observed flux after the flare stage.
During this stage, the injected energy is smaller than the kinetic energy of the GRB outflows.
The light curve is dominated by the initial energy of the jet.

As the magnetic inclination angle increases rapidly, the injected luminosity is powerful
enough to reshape the light curve. A break then appears in the solid curve.
Subsequently, as the increasing energy continues to be injected into the shock,
a rebrightening phase appears in the light curve.

As the angle stabilizes at $\pi/2$ rad, the solid curve overlaps with the dashed curve.
The flux of the evolving angle scenario matches that of the fixed-angle scenario.
The model with an evolving inclination angle satisfactorily accounts for the overall X-ray light
curve.

To further validate our model, we fit the multiband optical data of GRB 230414B.
As shown in Fig. \ref{fig3-OA}, the circles and squares correspond to the observational data in the R-band and I-band, respectively.
These data were obtained from the GRB Coordinates Network
\citep{Lu23,Raman23,Ghosh23,Mosk23,Male23,Zhu23,Belk23,Ror23,Quer23,Risin23}.
For our modeled R-band (solid curve) and I-band (dashed curve) synthetic light curves,
we incorporated dust extinction effects, using a Galactic extinction of $E(B-V) = 0.017$
\citep{Schl98,DAi23},
and host galaxy extinction assuming an LMC-like profile with $A_V = 0.4$ mag
\citep{Scha12}.

Fig. \ref{fig3-OA}  illustrates that our model fits most of the optical data for GRB 230414B.
The early-time flares (gray point) likely originate from ongoing central engine activity.
The optical light curves of GRB 230414B exhibit a rebrightening phase.
The observed rebrightening features in both bands result from the evolving inclination angle scenario.
The model is consistent with both the X-ray and optical afterglow data.

\section {Conclusions and discussion}

We investigated the rebrightening features in the afterglow
light curves of GRB 170822A and GRB 230414B.
We propose a new model to explain the rebrightenings.
In our scenario, a magnetar is born during the GRB explosion.
The nascent neutron star is prolate ellipsoid.
The magnetic inclination angle of the magnetar increases rapidly,
which leads to an increase in magnetic dipole radiation.

Our calculations demonstrate that as the inclination angle ($\alpha$) increases (e.g., from 0.01 to $\pi/2$ rad), the
energy loss rate (or energy injection luminosity) rises correspondingly.
This energy is injected into the GRB outflow, powering the afterglow emission through relativistic electron synchrotron radiation.
Using observational data from GRB 170822A and GRB 230414B, we demonstrate that afterglow rebrightenings
can be naturally reproduced within this framework.

Special jet structures or new components are not necessary to explain GRB rebrightening afterglows. The rebrightenings result from the increasing magnetic inclination angle.
No other structures or new components are required for the GRB explosions.
This scenario is also illustrated in our previous work \citep{Xu21}, in which a
decreasing inclination angle produces two-plateaus structures in GRB afterglows.
A single dynamical process can generate multiple features of GRB afterglows.

The proto-neutron star spins down according to Eq. (\ref{equ:L}). The magnetic dipole radiation
formula scales as $1+{\rm sin}^{2}\alpha$ \citep{Spit06}, unlike the vacuum case which scales as ${\rm sin}^{2}\alpha$.
The luminosity varies by at most a factor of approximately two between $\alpha$=0 and $90^\circ$. This factor is
significantly smaller than in the vacuum case. In our calculations, the energy loss rate $L$ increases by a factor of
approximately two (Fig. \ref{fig1-ELR}), implying that the rebrightening in the X-ray afterglow must be slight
(variation less than a factor of two).
Fig. \ref{fig2-XA} illustrates this limitation. Thus, our model is suitable only for slight rebrightenings.

The nascent neutron star in our model must possess a prolate ellipsoidal distortion.
Its deformation originates from a strong internal toroidal magnetic field.
Both electromagnetic and gravitational-wave emissions can influence the spin-down.
In our model, the magnetic ellipticity is approximately $10^{-7}$.  
The gravitational-wave emission luminosity is approximately $10^9$ times smaller than the
electromagnetic emission luminosity. The gravitational-wave emission can be neglected.
However, under more extreme magnetic field conditions, the gravitational-wave emission may play a key role.

Due to bulk viscous damping, the inclination angle of a prolate ellipsoid star generally increases to
$90^\circ$. We adopted and simplified the model from \citet{Land18}: the inclination
angle remains constant for the first $T_0/3$, then increases linearly from the initial angle
to the termination angle over the remaining $2T_0/3$.
The timescales of $T_0$ used in our model are longer than the typical bulk-viscosity timescales.
We emphasize that the timescale formula in \citet{Land18} is parameter-dependent.
Using the parameters listed in Table 1, we can recover the timescales.
The quantity $T_0$ is particularly sensitive to the temperature of the proto-neutron star,
which evolves rapidly at early time \citep{Pons99}.  
Consequently,  we obtain the value of $T_0$ only for a very fine choice of initial temperature, implying that the model requires a very narrow temperature range.
Our model is therefore particularly suitable for explaining early rebrightenings in X-ray afterglows.

In our model with an increasing magnetic inclination angle, the effects of the parameters on the afterglow light curves
are similar to those seen in models with fixed or decreasing inclination angles
\citep{Xu08,Xu09,Xu21}.
Although we do not present these specific results, our calculations show that the parameters related to
the magnetar, jet, and ISM significantly influence the light curves. Crucially, the shapes of features
such as the shallow decay, plateau, or rebrightening phases are primarily determined by the magnetar parameters.
All parameters satisfy the requirements of the external forward shock model.

Equation (\ref{equ:a}) shows that the evolution of the magnetic inclination angle is linear in our calculations. 
In reality, the evolution may be more complex 
\citep{Zana15}
, and we plan to consider nonlinear evolution in future work. 
We predict that such an evolution may lead to more complex features, such as oscillatory structures, 
in multiband afterglow light curves of GRBs. In our current model, breaks appear in the light curves when the injection
luminosity drops significantly. This suggests that some temporal structures in the afterglow light curves, 
such as shallow decays, plateaus, double plateaus, and rebrightenings, may share the same origin.

\begin{acknowledgements}
This work was supported by
the National Natural Science Foundation of China (Grant No. 11663003) and
the Science and Technology Research Project of Jiangxi Provincial Department of Education (Grant No. GJJ211110).
\end{acknowledgements}

\FloatBarrier 
\clearpage

\end{document}